\documentstyle[12pt]{article}
\begin{document}

\centerline{\large {\bf Parton Model From Field Theory via}}
\centerline{\large {\bf Light-Front Current Algebra:}}
\centerline{\large {\bf The Good, the Bad, and the
Terrible\footnote{
Talk Presented at the Fourth International Workshop on Light Cone
Quantization and Non-Perturbative Dynamics, Polana Zgorzelisko, Poland.}}}

\vskip.3in
\centerline{A. Harindranath$^*$}
\centerline{Saha Institute of Nuclear Physics}
\centerline{1/AF Bidhannagar, Calcutta 700064 India}

\vskip .6in

\centerline{\bf Abstract}

\vskip .1in

The emergence of parton model from field theory in the context of light-front
current algebra and naive canonical manipulations is reviewed. Shortcomings
of the naive canonical picture, especially concerning renormalization issues
are discussed. To illustrate the novel aspects of the renormalization
problem in light-front dynamics, the scaling behavior of different
components of currents under the dual power counting is stressed. It is noted
that the application of dual power counting to deep inelastic phenomena may
provide a simple intuitive understanding of twist.

\vskip .1in

\vskip .3in
{\bf Introduction}
\vskip .1in

The analysis and the resolution of renormalization problems associated with
Hamiltonians is an important area of study in light-front field theory.
Just like the Hamiltonian,
matrix elements of currents or products of currents are also directly related
to observables. Thus along with the study of renormalization problems
associated with light-front Hamiltonians we need the study of renormalization
problems associated with light-front currents.

Before we embark on the issues of renormalization of light-front current
matrix elements, however, it may be helpful to recall the canonical structure
of
light-front currents, their relevance to observables and the inferences
from canonical theory$^{1}$. It is useful to recognize from a physical
point of view the shortcomings of this picture so that we may be guided
in studying the problems of renormalization.

It is worthwhile to remember that one of the motivations for proposing
Quantum Chromodynamics (QCD) as the underlying theory of strong interactions
was indeed the structure of light-front current algebra$^{2}$.

In the following we review$^{1,2}$ the emergence of parton model from
canonical manipulations via light-front current algebra, mention
its shortcomings
from renormalization point of view, and briefly indicate how the dual power
counting on the light-front may be beneficial in addressing various
issues.

\vskip.3in
{\bf Fermionic Currents and Canonical Light-Front Commutators}
\vskip.1in

Borrowing from the Lagrangian formalism we may define the vector current
$ J^{\mu}(x) = {\bar \psi}(x) \gamma^{\mu} \psi(x) $
where $\psi(x)$ is the four-component Dirac field. (We will ignore the
internal flavor symmetry in this discussion).

In light-front variables, not all the four components of $\psi$ are dynamical.
It is customary to define projection operators $ \Lambda^{\pm} = {1 \over 4}
\gamma^{\mp} \gamma^{\pm}$ where $\gamma^{\pm} = \gamma^{0} \pm
\gamma^{3}$. Define $ \psi^{\pm} = \Lambda^{\pm} \psi$. In gauge theory
(QED or QCD), with the choice $A^{+}=0$, it follows from the equation of
motion that $\psi^{-}$ is constrained. Explicitly,
$$ \psi^{-}(x^- , x^\perp) = { - i \over 4} \int dy^- \epsilon(x^- - y^-)
[ i \alpha^\perp. \partial^\perp - g \alpha^\perp. A^\perp + \gamma^0 m]
\psi^{+}(y^- , x^\perp). $$
Thus the relation between $\psi^-$ and $\psi^+$ is nonlocal and as we
shall see the non-locality has far reaching consequences.

{}From the definition of current, we have,
$$J^{+} = 2 (\psi^{+})^{\dagger} \psi^{+}, \, \, \, J^{\perp}= (\psi^{+})^{
\dagger}\alpha^\perp \psi^{-} + (\psi^{-})^\dagger \alpha^\perp \psi^{+},
\, \, \, J^{-} = 2 (\psi^{-})^\dagger \psi^{-}. $$
Using the canonical commutation relation, $ \{ \psi^{+}(x),
(\psi^{+})^\dagger(y)\}_{x^+ = y^+} = \Lambda^{+} \delta^{3}(x-y)$, we get
$ [ J^{+}(x),J^{+}(y) ]_{x^+ = y^+} =0$. To compute
$ [ J^{+}(x), J^{-}(y)]$
we need the equation of constraint and hence the
equation of motion. We have
$$ [ J^{+}(x), J^{-}(y)]_{x^+=y^+} = \partial^{+}_{x} \Big \{
  -{ 1 \over 2} \epsilon (x^- - y^-) \delta^{2}(x^\perp - y^\perp)
                     {\bar \psi}(x) \gamma^- \psi(y) \Big \} $$
$$ \, \, \, \, \, + \partial^{i}_{x} \Big \{
{ 1 \over 2} \epsilon(x^- - y^-) \delta^{2}(x^\perp - y^\perp)
\big [ {\bar \psi}(x)\gamma^{i} \psi(y) + i \epsilon^{ij}
       {\bar \psi}(x)\gamma^{j} \gamma^{5} \psi(y) \big ]
\Big \}. $$

Thus bilocal vector and axial vector currents emerge canonically. It is
important to note that the non-locality is only in the longitudinal ($x^-$)
direction.

For future use define
$$ {\cal J}^{\mu}(x|y) = { 1 \over 2}\big(J^{\mu}(x|y) + J^{\mu}(y|x)\big),
\, \, \, {\bar {\cal J}}^{\mu}(x|y) = { 1 \over 2i}
\big(J^{\mu}(x|y) - J^{\mu}(y|x)\big),$$
$$<P| {\cal J}^{\mu}(y|0) |P> = P^{\mu} V_{1}(y^2, P.y) + y^{\mu}
V_{2} (y^2, P.y), $$
$$<P| {\bar{\cal J}}^{\mu}(y|0) |P> = P^{\mu} {\bar V_{1}}(y^2, P.y)
 + y^{\mu} {\bar V_{2}} (y^2, P.y). $$
Note that $V(y^2,P.y) \rightarrow V({1 \over 2} P^+ y^-) \, \,
= V(\eta) $ since
$y^2=0$, and $P.y={1 \over 2} P^+ y^- =\eta$ at $y^+=0, y^\perp=0$.

\vskip.3in
{\bf Scaling Function as Fourier Transform of Bilocal Matrix Element}
\vskip.1in

The hadron tensor relevant for spin-averaged electron-nucleon scattering
is given by
$$ W^{\mu \rho} = {1 \over 4 \pi} \int d^4y \,e^{iq.y}\, <P| [ J^{\mu}(y),
J^{\rho}(0)]|P> $$
$$ \, \, \, \, \, =[-g^{\mu \rho} + { q^{\mu} q^{\rho} \over q^2}]W_1
+ (P^\mu - {P.q q^\mu \over q^2})(P^\rho - {P.q q^\rho \over q^2})W_2. $$
Consider forward virtual Compton scattering amplitude
$$ T^{\mu \rho}(P,q) = i \int d^4y \, \theta(y^+) \, e^{i q.y}
<P| [ J^{\mu}(y), J^{\rho}(0)]|P>. $$
We also have
$$ W^{\mu \rho}(P,q) = { 1 \over 2 \pi}
 Im \, T^{\mu \rho}(P,q) . $$
Write a fixed $q^2$ dispersion relation
$$ T^{\mu \rho} = \int {d \nu' \over \nu' - \nu} W^{\mu \rho} \, \, \,
{\rm where}\, \, \nu = P.q \, \, .\eqno(1)$$
Consider Bjorken-Johnson-Low limit of Compton amplitude
$$ Limit_{q^- \rightarrow \infty}\,  T^{\mu \rho}
= - { 1 \over q^-} \int dy^- d^2 y^\perp
e^{i({q^+y^- \over 2}- q^\perp . y^\perp)}
<P| [ J^{\mu}(y), J^{\rho}(0)]_{y^+=0}|P>. $$
Using the integral representation for the antisymmetric step function

$$ \epsilon(x^-) = - {i \over \pi} \int {dq' \over q'} e^{{i \over 2}
q'x^-}$$
and taking the $q^- \rightarrow \infty $ of eq. (1), take absorptive part
on both sides and compare coefficients.
{}From "$+-$" component,
$$ Limit_{q^- \rightarrow \infty}\, \nu W_2(q^2, \nu) = F_2(x) $$ with
$x = {-q^2 \over 2 \nu}$ and
${F_2(x) \over x} =  { i \over 2 \pi} \int d \eta e^{-i \eta x }
{\bar V}_1(\eta)$.
Thus scaling function is the Fourier transform of the bilocal matrix element.
Just as matrix elements of local currents are measured in elastic scattering,
deep inelastic scattering measures matrix elements of bilocal currents.
{}From "++" component we get $Limit_{q^- \rightarrow \infty} W_2, W_L =0$,
where $ W_L = W_1 + {(P.q)^2 \over q^2} W_2$.

Making a Fock space expansion for $|P>$, i.e.,
$$ |P> = \int \, \phi(k_1) b^{\dagger}(k_1) d^{\dagger}(P-k_1) |0> \,
+ ...  . $$
the parton picture with probabilistic interpretation emerges:
$$ {F_2(x) \over x } = \int d^2 k^\perp |\phi(x,k^\perp)|^2 + ...  . $$

\vskip.3in
{\bf Trouble with Canonical Picture: Renormalization Aspects}
\vskip.1in
In the naive canonical manipulations (even though they have lead to an
intuitive physical picture of scaling) renormalization effects are completely
ignored. The structure function we obtained has no dependence on a mass scale
whereas in the real world we do need a scale dependence. Once renormalization
effects are taken in to account, we encounter divergent loop integrals when
loop momenta also tend to infinity. Thus $q^- \rightarrow \infty$ limit is
valid a priori only for a cutoff theory. So the question remains whether
the intuitive parton based picture still survives after renormalization
effects are taken into account.

\vskip.3in
{\bf Dual Scaling Analysis, Light-front Power Counting, Consequences}
\vskip.1in
To start tackling the renormalization problem which is forced upon us from
physical considerations, let us begin with light-front canonical reasonings.
The starting point of renormalization analysis is the study of behavior of
operators under scale transformations. In light-front dynamics we consider
separate scaling analysis$^{3}$ in the longitudinal ($x^-$) coordinate and
transverse ($x^\perp$) coordinate. For the fermion field operators we have
$ \psi^+ \sim { 1 \over \sqrt{x^-}} { 1 \over x^\perp} $ and
$ \psi^- \sim { \sqrt{x^-} \over (x^\perp)^2}$. Thus $J^+
\sim { 1 \over x^-} { 1 \over (x^\perp)^2}$ and canonically $J^+$
has a unique
scaling behavior whether or not masses are present. On the other hand
$J^\perp \sim { 1 \over (x^\perp)^3}$ and $J^- \sim { x^- \over
(x^\perp)^4}$. When masses are present, $J^\perp$ and $J^-$ have no unique
transverse scaling behavior and only dimensional analysis applies in the
transverse coordinate.

First let us recall the consequences for scale breaking which follow from
canonical reasonings. In light-front dynamics longitudinal scale
transformation corresponds to longitudinal boost transformation and hence
longitudinal scale invariance is a Lorentz symmetry of the theory. Hence
canonical reasonings indicate that the longitudinal scale invariance cannot
be broken by masses or renormalization process. This implies that a mass
scale can get generated only through transverse divergences. This inference
is corroborated in perturbation theory; for example, the standard asymptotic
freedom result in light-front QCD arises through transverse momentum
divergences.

Consider deep inelastic process where relevant distance scales are
short transverse separations and medium to large longitudinal separations.
The dual power counting is ideally suited to study this phenomena.
The behavior of bilocal matrix elements for large $y^-$ determines the
small $x$ behavior of structure functions. On the other hand, power
correction to scaling is determined by the scaling behavior of operator
product of currents under transverse scale transformations.

To study power corrections to scaling one can classify operators on the basis
of their transverse mass dimension. For example, $J^+$ has transverse mass
dimension 2. According to the terminology of Gell-Mann and Fritzsch $+, \,
\perp, \, $ and $-$ components correspond to good, bad and terrible operators
respectively. For good operators we notice that twist and
transverse mass dimension coincide. However, by the same token, bad and
terrible operators correspond to higher twist! From different considerations
similar conclusions have been arrived at before
by parton theorists$^{4}$.

\vskip.3in
{\bf Acknowledgement}
\vskip.1in
The author acknowledges helpful conversations with the members
of the Theory Group, SINP. He is also grateful to the participants of the
Polana Zgorzelisko Workshop, especially Prof. Leonard Susskind for many
illuminating
discussions. The author thanks Prof. Jozef Namys{\l}owski for his kind
hospitality in Warsaw, Tatra mountains and places in between.

\vskip.3in
{\bf References}
\vskip.1in
$^*$  e-mail: hari@tnp.saha.ernet.in
\vskip.1in
1. J. M. Cornwall and R. Jackiw, Phys. Rev. {\bf D4}, 367 (1971); D. A. Dicus,
R. Jackiw and V. L. Teplitz, Phys. Rev. {\bf D4}, 1733 (1971); T. M. Yan,
Phys. Rev. {\bf D7}, 1760 (1973); V. De Alfaro, S. Fubini, G. Furlan, and
G. Rossetti, {\it Currents in Hadron Physics}, (North-Holland, Amsterdam,
1973).
\vskip.1in
2. See, for example, H. Fritzsch and M. Gell-Mann, Light Cone Current Algebra,
in Proceedings of the International Conference on Duality and Symmetry in
Hadron Physics, edited by E. Gotsman, The Weizman Press of Israel, Jerusalem
(1971); H. Fritzsch, M. Gell-Mann and H. Leutwyler, Phys. Lett. {\bf 47B},
365 (1973).
\vskip .1in
3. K. G. Wilson, {\it Light-front QCD}, OSU internal report, (1990);
K. G. Wilson, T. Walhout, A. Harindranath, W.-M. Zhang, R.J. Perry, and
St. D. Glazek, Phys. Rev. {\bf D49}, 6720 (1994).
\vskip.1in
4. J. C. Collins and D. E. Soper, Nucl. Phys. {\bf B194}, 445 (1982);
R. L. Jaffe and X. Ji, Nucl. Phys. {\bf B375}, 527 (1992).

\end{document}